\documentstyle[aps,epsf]{revtex}

\begin{document}
\draft

\twocolumn[
\hsize\textwidth\columnwidth\hsize\csname@twocolumnfalse\endcsname\title
{A simple model of price formation}
\author{K. Sznajd--Weron$^{\dag}$
        \and\ R. Weron$^{\ddag}$}
\address{$^{\dag}$ Institute of Theoretical Physics, University of Wroc{\l}aw, \\
         pl. Maxa Borna 9, 50-204 Wroc{\l}aw, Poland \\
         $^{\ddag}$ Hugo Steinhaus Center, Wroc{\l}aw University of Technology,\\
         Wyspianskiego 27, 50-370 Wroc{\l}aw, Poland}
\maketitle

\begin{abstract}
A simple Ising spin model which can describe the mechanism of price formation
in financial markets is proposed. In contrast to other agent-based models, the influence
does not flow inward from the surrounding neighbors to the center site, but spreads
outward from the center to the neighbors. The model thus describes the spread of opinions
among traders.
It is shown via standard Monte Carlo simulations that very simple rules lead to dynamics
that duplicate those of asset prices.
\end{abstract}

]

The Ising spin system is one of the most frequently used models of statistical mechanics.
Its simplicity (binary variables) makes it appealing to researchers from other branches
of science including biology \cite{dh91}, sociology \cite{sh00} and economy
\cite{smr99,cggs99,cs99,cb00,cmz00,ez00}.
It is rather obvious that Ising-type models cannot explain origins of very complicated
phenomena observed in complex systems. However, it is believed that these kind of models
can describe some universal behavior.

Recently, an Ising spin model which can describe the mechanism of making a decision
in a closed community was proposed \cite{s-ws00}. In spite of simple rules the model
exhibited complicated dynamics in one and two \cite{sso00} dimensions.
In contrast to usual majority rules \cite{adler91}, in this model the influence was
spreading outward from the center. This idea seemed appealing and we adapted it to model
financial markets. We introduced new dynamic rules describing the behavior of two types
of market players: trend followers and fundamentalists. The obtained results were
astonishing -- the properties of simulated price trajectories duplicated those of
analyzed historic data sets. Three simple rules led to a fat-tailed distribution
of returns, long-term dependence in volatility and no dependence in returns themselves.

We strongly believe that this simple and parameter free model is a good first approximation
of a number of real financial markets. But before we introduce our model we review some of
the stylized facts about price formation in the financial markets.

{\bf Stylized facts.}
---
Adequate analysis of financial data relies on an explicit definition of the variables
under study. Among others these include the price and the change of price.
In our studies the price $x_t$ is the daily closing price for a given asset.
The change of price $r_t$ at time $t$ is defined as $r_t = \log x_{t+1} - \log x_t$.
In fact, this is the change of the logarithmic price and is often referred
to as return. The change of price, rather than the price itself, is the
variable of interest for traders (and researchers as well).

{\it Fat-tailed distribution of returns.}
---
The variety of opinions about the distributions of asset returns and their
generating processes is wide. Some authors claim the distributions to be close to
Paretian stable \cite{rm00}, some to generalized hyperbolic \cite{bg74,ek95},
and some reject any single distribution \cite{c-rb-h82,ms94}.

Instead of looking at the central part of the distribution, an alternative way
is to look at the tails. Most types of distributions can be classified into three
categories:
1$^{\circ}$ thin-tailed -- for which all moments exist and whose density
function decays exponentially in the tails;
2$^{\circ}$ fat-tailed -- whose density function decays in a power-law fashion;
3$^{\circ}$ bounded -- which have no tails.

Virtually all quantitative analysts suggest that asset returns fall into the second
category. If we plot returns against time we can notice many more outlying (away from
the mean) observations than for white noise. This phenomenon can be seen even better
on normal probability plots, where the cumulative distribution function (CDF) is
drawn on the scale of the cumulative Gaussian distribution function.
Normal distributions have the form of a straight line in this representation,
which is approximately the case for the distribution of weekly or monthly returns.
However, distributions of daily and higher-frequency returns are distinctly fat-tailed
\cite{olsen90}. This can be easily seen in the top panels of Fig. 1, where daily
returns of the DJIA index for the period Jan. 2nd, 1990 -- Dec. 30th, 1999, are presented.

{\it Clustering and dependence.}
---
Despite the wishes of many researchers asset returns cannot be modeled
adequately by series of iid (independent and identically distributed) realizations
of a random variable described by a certain fat-tailed distribution.
This is caused by the fact that financial time series depend on the evolution
of a large number of strongly interacting systems and belong to the class of
complex evolving systems. As a result, if we plot returns against time we can
observe the non-stationarity (heteroscedasticity) of the process in the form of
clusters, i.e. periods during which the volatility (measured by standard deviation
or the equivalent $l^1$ norm \cite{olsen97}) of the process is much higher than usual,
see the top-left panel of Fig. 1. Thus it is natural to expect dependence in asset returns.
Fortunately, there are many methods to quantify dependence. The direct method consists
in plotting the autocorrelation function:
${\bf acf}(r,k) = {\sum_{t=k+1}^{N} (r_t-\bar{r})(r_{t-k}-\bar{r})} /
 {\sum_{t=1}^{N} (r_t-\bar{r})^2} $,
where $N$ is the sample length and $\bar{r} = \frac{1}{N} \sum_{t=1}^{N} r_t$,
for different time lags $k$. For most financial data autocorrelation of returns dies out
(or more precisely: falls into the confidence interval of Gaussian random walk) after at
most a few days and long-term autocorrelations are found only for squared or absolute
value of returns \cite{olsen97,ww98,w00}, see bottom panels of Fig. 1. Recall that for
Brownian motion -- the classical model of price fluctuations \cite{ww98,bachelier00}
-- autocorrelations of $r_t$, $r_t^2$ and $|r_t|$ are not significant for lags greater or
equal to one.

\begin{figure}[htbp]
\centerline{\epsfxsize=8.5cm \epsfbox{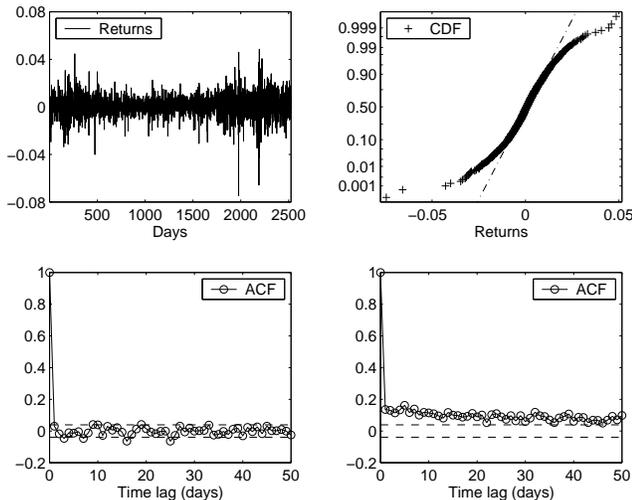}}
\caption{Daily returns of the Dow Jones Industrial Average index during the last decade
({\it top left}), normal probability plot of DJIA returns ({\it top right}),
lagged autocorrelation function of DJIA daily returns ({\it bottom left}), and
lagged autocorrelation function of absolute value of DJIA daily returns ({\it bottom right}).
Dashed horizontal lines represent the 95\% confidence interval of a Gaussian random walk.}
\end{figure}

Another way to examine the dependence structure is the power spectrum analysis, also known
as the "frequency domain analysis". One of the most often used techniques was proposed by
Geweke and Porter-Hudak \cite{gp-h83} (GPH) and is based on observations of the slope of
the spectral density function of a fractionally integrated series around the angular
frequency $\omega = 0$. A simple linear regression of the periodogram $I_n$ (a sample
analogue of the spectral density) at low Fourier frequencies $\omega_k$:
$\log\{I_n(\omega_k)\} = a - \hat{d} \log\{4 \sin^2 (\omega_k/2)\} + \epsilon_k$
yields the differencing parameter $d=H-0.5$ through the relation $d = \hat{d}$.
The GPH estimate of the Hurst exponent $H$ has well known asymptotic properties and allows
for construction of confidence intervals \cite{gp-h83,weron01}.

Yet another method is the Hurst R/S analysis \cite{hurst51,mw69} or its
"younger sister" -- the Detrended Fluctuation Analysis (DFA) \cite{boston94}. Both
methods are based on a similar algorithm, which begins with dividing the time series
(of returns) into boxes of equal length and normalizing the data in each box by
subtracting the sample mean (R/S) or a linear trend (DFA). Next some sort of
volatility statistics is calculated (rescaled range or mean square fluctuation,
respectively) and plotted against box size on a double-logarithmic paper. Linear
regression yields the Hurst exponent $H$, whose value can lie in one of the three regimes:
1$^{\circ}$ $H>0.5$ -- persistent time series (strict long memory),
2$^{\circ}$ $H=0.5$ -- random walk or a short-memory process,
3$^{\circ}$ $H<0.5$ -- anti-persistent (or mean-reverting) time series.
Unfortunately, no asymptotic distribution theory has been derived for the R/S and DFA
statistics so far. However, it is possible to estimate confidence intervals based on
Monte Carlo simulations \cite{weron01}.

{\bf The model.}
---
Recently a simple model for opinion evolution in a closed community was proposed
\cite{s-ws00,sso00}. In this model (called USDF) the community is represented by a horizontal
chain of Ising spins which are either up or down. A pair of parallel neighbors forces
its two neighbors to have the same orientation (in random sequential updating), while
for an antiparallel pair, the left neighbor takes the orientation of the right part
of the pair, and the right neighbor follows the left part of the pair.

In contrast to usual majority rules \cite{adler91}, in the USDF model the influence does
not flow inward from the surrounding neighbors to the center site, but spreads outward
from the center to the neighbors. The model thus describes the spread of opinions.
The dynamic rules lead to three steady states: two ferromagnetic (all spins up or all
spins down) and one antiferromagnetic (an up-spin is followed by a down-spin, which is
again followed by an up-spin, etc.).

In this paper we modify the model to simulate price formation in a financial market.
The spins are interpreted as market participants' attitude. An up-spin ($S_i=1$)
represents a trader who is bullish and places buy orders, whereas a down-spin ($S_i=-1$)
represents a trader who is bearish and places sell orders. In our model the first
dynamic rule of the USDF model remains unchanged, i.e.
%\begin{itemize}
%\item
{\it if $S_i S_{i+1}=1$ then $S_{i-1}$ and $S_{i+2}$ take the direction of the pair (i,i+1)}.
%\end{itemize}
This can be justified by the fact that a lot of market participants are trend followers
and place their orders on the basis of a local guru's opinion.
However, the second dynamic rule of the USDF model has to be changed to incorporate
the fact that the absence of a local guru (two neighboring spins are in different
directions) causes market participants to act randomly rather than make the opposite
decision to his neighbor:
%\begin{itemize}
%\item
{\it if $S_i S_{i+1}=-1$ then $S_{i-1}$ and $S_{i+2}$ take one of the two directions
at random}.
%\end{itemize}

Such a model has two stable states (both ferromagnetic), which is not very realistic for
a financial market. Fortunately, trend followers are not the only participants of the
market \cite{bps97}. There are also fundamentalists -- players that know much more about
the system and have a strategy (or perhaps we should call them "rationalists").
To make things simple, in our model we introduce one fundamentalist. He knows exactly
what is the current difference between demand and supply in the whole system.
If supply is greater than demand he places buy orders, if lower -- sell orders.

It is not clear {\it a priori} how to define the price in a market. The only obvious
requirement is, that the price should go up, when there is more demand than supply,
and vice versa. For simplicity, we define the price $x_t$ in our model as the normalized
difference between demand and supply (magnetization):
$x_t = \frac{1}{N}\sum_{i=1}^N S_i(t)$,
where $N$ is the system size. Note that $x_t\in [-1, 1]$, so $|x_t|$ can be
treated as probability. Now we can formulate the third rule of our model:
%\begin{itemize}
%\item
{\it the fundamentalist will buy (i.e. take value $1$) at time $t$ with probability
$|x_t|$ if $x_t<0$ and sell (i.e. take value $-1$) with probability $|x_t|$ if $x_t>0$}.
%\end{itemize}

The third rule means that if the system will be close to the stable state "all up",
the fundamentalist will place sell orders with probability close to one (in the
limiting state exactly with probability one) and start to reverse the system.
So the price will start to fall. On the contrary, when $r$ will be close to $-1$,
the fundamentalist will place buy orders (take the value $1$) and the price will
start to grow. This means that ferromagnetic states will not be stable states anymore.

{\bf Results.}
---
To investigate our model we perform a standard Monte Carlo simulation with random updating.
We consider a chain of $N$ Ising spins with free boundary conditions. We were usually
taking $N=1000$, but we have done simulations for $N=10000$ as well. We start from a totally
random initial state, i.e. to each site of the chain we assign an arrow with a randomly
chosen direction: up or down (Ising spin).

\begin{figure}[htbp]
\centerline{\epsfxsize=8.5cm \epsfbox{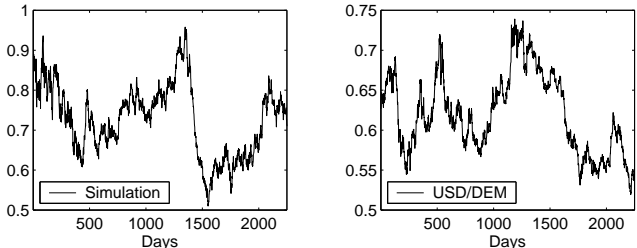}}
\caption{A typical path of the simulated price process $x_t$ and the USD/DEM exchange rate,
respectively.}
\end{figure}

In our simulations each Monte Carlo step (MCS) represents one trading hour; eight steps
constitute one trading day. We typically simulate 20000 MCSs, which corresponds to
2500 trading days or roughly 10 years. We chose such a period of time, because in this
paper we compare the empirical results with historical data sets (two FX rates and two
stock indices) of about the same size:
2245 daily quotations of the dollar-mark (USD/DEM) exchange rate for the period Aug. 9th,
1990 -- Aug. 20th, 1999 (see Fig. 2 and Table 1),
2809 daily quotations of the yen-dollar (JPY/USD) exchange rate for the period Jan. 2nd,
1990 -- Feb. 28th, 2001 (see Table 1),
2527 daily quotations of the Dow Jones Industrial Average (DJIA) index for the period
Jan. 2nd, 1990 -- Dec. 30th, 1999 (see Fig. 1),
and 1561 daily quotations of the WIG20 Warsaw Stock Exchange index (based on 20 blue chip stocks
from the Polish capital market) for the period Jan. 2nd, 1995 -- Mar. 30th, 2001 (see Table 1).

The returns $r_t$ are obtained from the simulated price curve $x_t$ (see Fig. 2)
after it is shifted (incremented by one) to make it positive. Alternatively we
could have defined the up-spin to be equal to two and the down-spin -- to zero. However,
this would have made the calculations more difficult and the description of the model
less appealing.

In Figure 3 we present daily returns and normal probability plots for the simulated
(left panels) and USD/DEM exchange rate (right panels) time series. Without prior
knowledge as to the magnitude of historical returns it is impossible to judge which
process is real and which is a fraud. The same is true for the simulated and the DJIA
returns of Fig. 1.

\begin{figure}[htbp]
\centerline{\epsfxsize=8.5cm \epsfbox{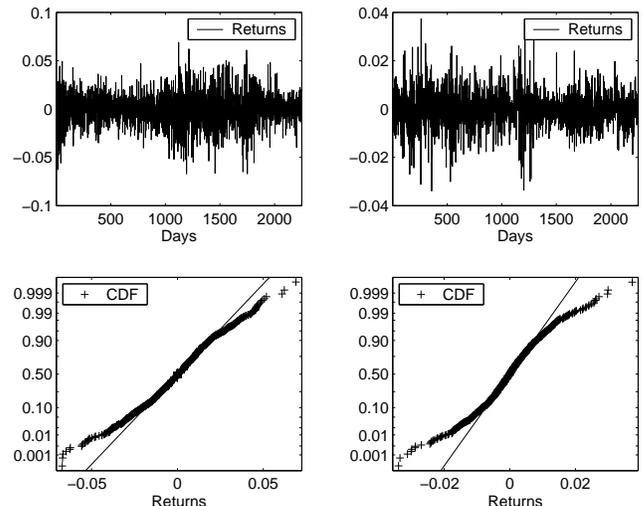}}
\caption{Returns of the simulated price process $x_t$ and daily returns of
the USD/DEM exchange rate during the last decade, respectively ({\it top panels}).
Normal probability plots of $r_t$ and USD/DEM returns, respectively,
clearly showing the fat tails of price returns distributions ({\it bottom panels}).}
\end{figure}

In Figure 4 we present the lagged autocorrelation study for the simulated
(left panels) and USD/DEM exchange rate (right panels) time series. Again
the properties of the simulated price process duplicate those of historical data.
The lagged autocorrelation of returns falls into the confidence interval of Gaussian
random walk immediately, like for the dollar-mark exchange rate. The same plot
for the DJIA returns shows a bit more dependence.

This can be seen also in Table 1, where values and significance of the R/S, DFA and GPH
statistics for the simulated and four historical data sets are presented. In all but one
case (DFA for DJIA) the Hurst exponents of daily returns are insignificantly different
from those of white noise. On the other hand, the Hurst exponents of the absolute value
of daily returns are persistent in all cases, with the results being significant even
at the two-sided 99\% level. Moreover, the estimates of $H$ from the simulation are
indistinguishable from those of real market data.

\begin{figure}[htbp]
\centerline{\epsfxsize=8.5cm \epsfbox{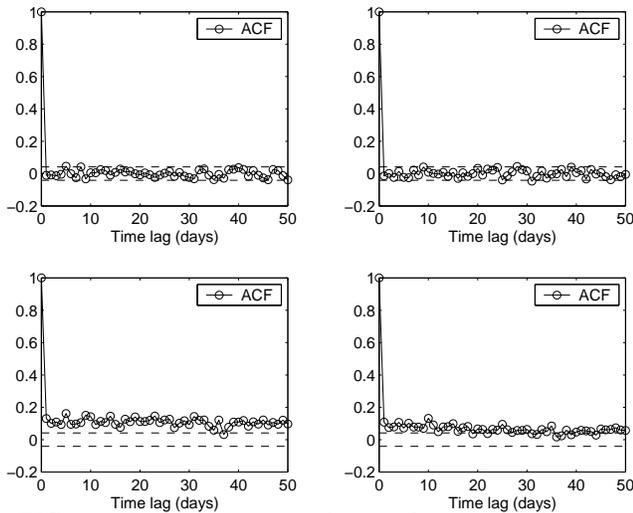}}
\caption{Lagged autocorrelation functions of $r_t$ and USD/DEM returns,
respectively ({\it top panels}). Lagged autocorrelation functions of absolute value
for the same data sets ({\it bottom panels}).
Dashed horizontal lines represent the 95\% confidence interval of a Gaussian random walk.}
\end{figure}

The presented empirical analysis clearly shows that three simple rules of our model lead
to a fat-tailed distribution of returns, long-term dependence in volatility and no dependence
in returns themselves as observed for market data. Thus we may conclude that this simple model
is a good first approximation of a number of real financial markets.

We gratefully acknowledge critical comments of an anonymous referee, which led to
a substantial improvement of the paper.

\begin{table}[htbp]
\caption{Estimates of the Hurst exponent $H$ for simulated and market data.
}
\begin{center}
\begin{tabular}{llll}
           & & Method & \\
Data       & R/S-AL & DFA & GPH \\
\hline
 & \multicolumn{3}{c}{\it Returns} \\
Simulation       & 0.5270 & 0.4666 & 0.3653 \\
USD/DEM          & 0.5127 & 0.5115 & 0.6154 \\
JPY/USD          & 0.5353 & 0.5303 & 0.5790 \\
DJIA             & 0.4585 & 0.4195$^{**}$ & 0.3560 \\
WIG20            & 0.5030 & 0.4981 & 0.4604 \\
\hline
 & \multicolumn{3}{c}{\it Absolute value of returns} \\
Simulation       & 0.8940$^{***}$ & 0.9335$^{***}$ & 0.8931$^{***}$ \\
USD/DEM          & 0.7751$^{***}$ & 0.8406$^{***}$ & 0.8761$^{***}$ \\
JPY/USD          & 0.8576$^{***}$ & 0.9529$^{***}$ & 0.9287$^{***}$ \\
DJIA             & 0.7838$^{***}$ & 0.9080$^{***}$ & 0.8357$^{***}$ \\
WIG20            & 0.9103$^{***}$ & 0.9494$^{***}$ & 0.8262$^{***}$ \\
\end{tabular}
\end{center}
{\small
$^*$, $^{**}$ and $^{***}$ denote significance at the (two-sided) 90\%, 95\% and 99\% level,
respectively. For the R/S-AL and DFA statistics inference is based on empirical
Monte Carlo results of Weron \cite{weron01}, whereas for the GPH statistics --
on asymptotic distribution of the estimate of $H$ \cite{gp-h83}.}
\end{table}

\end{document}